\documentclass[osajnl,twocolumn,showpacs,superscriptaddress,showkeys,10pt]{revtex4-1} 

\usepackage[T1]{fontenc}
\newcommand{\e}{\mathrm{e}}
\newcommand{\ic}{i}
\newcommand{\B}{\boldsymbol}
\newcommand{\tens}[1]{\B{#1}}
\newcommand{\re}{\mathfrak{Re}}
\newcommand{\im}{\mathfrak{Im}}
\newcommand*\conj[1]{
  \vbox{
  \hrule height 0.3pt
  \kern0.5ex
  \hbox{
  \kern-0.4em
  \ifmmode#1\else\ensuremath{#1}\fi
   \kern-0.em
  }
 } 
}

\newcommand{\rot}{\bm{\mathrm{\nabla\times}}}

\newcommand{\Mop}{\mathcal{M}}
\newcommand{\Lop}{\mathcal{L}}
\newcommand{\ddroit}{\mathrm{d}}
\newcommand{\source}{\bm{\mathcal{S}}}

\newcommand{\bra}{\ensuremath{\left\langle}}
\newcommand{\ket}{\ensuremath{\right\rangle}}
\newcommand{\mb}{\ensuremath{\,\middle| \,}}

\usepackage{graphicx}
 \usepackage[utf8x]{inputenc}
\usepackage{amssymb,amsmath}
 \usepackage{siunitx}
\usepackage[overload]{empheq}
\usepackage{subfigure}
\usepackage{numprint}
\usepackage{bm}
\usepackage{pgfplots}

 \usepackage{tikz}
  \usetikzlibrary{%
      decorations.pathreplacing,%
      decorations.pathmorphing,%
  shapes%
  }
\PassOptionsToPackage{dvipsnames}{xcolor}
	\RequirePackage{xcolor} 
\definecolor{halfgray}{gray}{0.55} 
\definecolor{webgreen}{rgb}{0,.5,0}
\definecolor{webbrown}{rgb}{.6,0,0}

  \definecolor{lgreen} {RGB}{180,210,100}
  \definecolor{dblue}  {RGB}{20,66,129}
  \definecolor{lred}   {RGB}{220,0,0}
  \definecolor{nred}   {RGB}{224,0,0}
  \definecolor{norange}{RGB}{210,129,34}
  \definecolor{nyellow}{RGB}{255,221,0}
  \definecolor{ngreen} {RGB}{98,158,31}
  \definecolor{dgreen} {RGB}{78,138,21}
  \definecolor{nblue}  {RGB}{28,130,185}
  \definecolor{jblue}  {RGB}{20,50,100}
  \usepackage[active]{srcltx}

\usepackage{lipsum}
  
\begin{document}


\title{Resonant metamaterial absorbers for infrared spectral filtering: quasimodal analysis, design, fabrication and characterization}


\author{Benjamin Vial}
\email[]{benjamin.vial@fresnel.fr}
\affiliation{Centrale Marseille, Aix Marseille Université, CNRS, Institut Fresnel, UMR 7249, 13013 Marseille, France}
\affiliation{Silios Technologies, ZI Peynier-Rousset, rue Gaston Imbert Prolong\'ee, 13790 Peynier, France}
\author{Guillaume Dem\'esy}
\author{Frédéric Zolla}
\author{André Nicolet}
\author{Mireille Commandré}
\author{Christophe Hecquet}
\author{Thomas Begou}
\affiliation{Centrale Marseille, Aix Marseille Université, CNRS, Institut Fresnel, UMR 7249, 13013 Marseille, France}

\author{St\'ephane Tisserand}
\author{Sophie Gautier}
\author{Vincent Sauget}
\affiliation{Silios Technologies, ZI Peynier-Rousset, rue Gaston Imbert Prolong\'ee, 13790 Peynier, France}

\date{\today}

\begin{abstract}
We present a modal analysis of metal-insulator-metal (MIM) based metamaterials in the far infrared region. 
These structures can be used as resonant reflection bandcut spectral filters that are independent of the polarization and direction of incidence 
 because of the excitation of quasimodes (modes associated with a 
complex frequency) leading to quasi-total absorption.
We fabricated large area samples made of chromium nanorod gratings on top of Si/Cr layers deposited on silicon substrate 
 and measurements by Fourier Transform spectrophotometry show good 
 agreement with finite element simulations. A quasimodal expansion method is developed 
 to obtain a reduced order model that fits very well full wave simulations and that highlights excitation conditions of the modes.
\end{abstract}

\pacs{}
\keywords{metamaterials, infrared, absorbers, filtering, quasimodes, finite element method}

\maketitle

\section{Introduction}

Structuration of metallic surfaces with typical size smaller than the wavelength 
can lead to spectacular resonant effects. More than one century ago, anomalies in reflection of metallic gratings have been discovered by Wood \cite{wood1902rcu}, 
and substantial pioneering work \cite{Hutley1976431,Hessel1965} have highlighted the role of surface plasmons polaritons in the 
anomalous reflection in mono and bi-periodic gratings. These resonances can be used to fashion various reflection and transmission spectra. 
In particular, total absorption phenomena in different metamaterial 
type \cite{Teperik2008,Landy2008,Bonod2008,Tao2008,Hao2010} from the micro wave to optical regime, have recently attracted a lot of interest 
because of their potential application in sensing \cite{Liu2010}, tunable frequency selective microbolometers \cite{Maier2009,Maier2010} 
or solar cells \cite{Teperik2008}. One family of metamaterial have been extensively studied which is 
based on Metal-Insulator-Metal (MIM)configuration \cite{Hao2010,Aydin2011,Bouchon2012,Hao2011}, because they 
can lead to polarization and angle independent resonant perfect absorption. This is the kind of structures we study 
both numerically and experimentally
 in this paper with the aim of using them as bandcut reflection filters in the infrared that can be tuned by 
adjusting the periodicity of the grating.\\
Besides the calculation of diffraction efficiencies and absorption spectra, our approach to 
study the resonant phenomena in such metamaterials is to compute the eigenmodes and eigenfrequencies of such 
open electromagnetic systems. The study of poles and zeros of the scattering operator \cite{Popov1986,Neviere1995} and 
of their associated leaky modes leads to significant insights into the properties of
metamaterials \cite{Tikhodeev2002,FEHREMBACH2003,Lalanne2006,Grigoriev2013} and eases the conception diverse optical devices
 \cite{fehrembach,sentenac2005atr,Ding1,Ding2004,Ding2004a} because it provides a simple picture of 
the resonant processes at stake. From the resolution of a spectral problem, one obtains complex eigenfrequencies.
 The real part is the resonant frequency and the imaginary part the bandwidth. Resonant 
scattering is expected when shining light with frequency around the resonant frequency. We report here a numerical 
spectral analysis of MIM arrays, that allows us to optimize parameters 
for infrared reflection bandcut filters. The spectral position od the reflection dip can be adjusted by varying the periodicity of the grating. 
Large area samples with different periods have been fabricated and characterized by FTIR spectroscopy, and measured normal incidence 
reflection spectra agree well with the numerical predictions of both calculated reflection spectra and complex eigenvalues. 
Moreover, the high angular tolerance of the filters is demonstrated experimentally and numerically.\\
The eigenvectors and eigenvalues are 
intrinsic properties of the studied system that depends onto the opto-geometrical parameters but are in essence 
independent of the incident parameters. 
Our main contribution is to provide a 
systematic method to characterize the excitation of a given mode. By expanding 
the scattered field onto the eigenmode basis, we can compute the coupling coefficient that characterizes the strength of the 
interaction of incident light with a mode. This method is illustrated in the case of a MIM array, 
showing the resonant nature of the reflection dip and providing a reduced-order model with two degenerate leaky modes 
that fits very well full wave finite elements calculation.

\section{Setup of the problem and theoretical background}\label{setup}

\begin{figure}[htbp!]
\begin{center}
\begin{tabular}{c}
\subfigure[\label{schema1}]{\includegraphics[width=0.68\columnwidth]{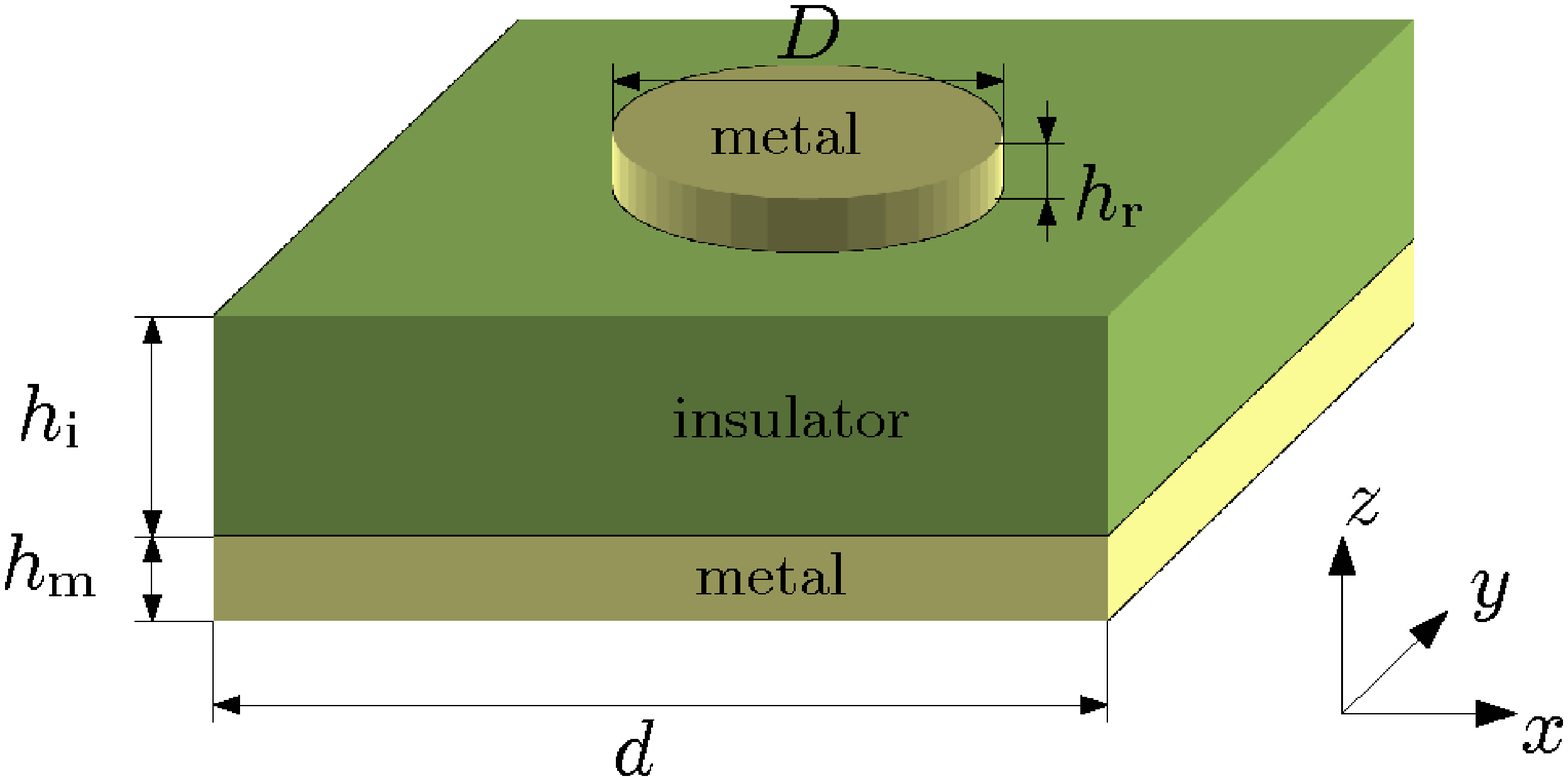}}
\subfigure[\label{fab}]{  \includegraphics[width=0.28\columnwidth]{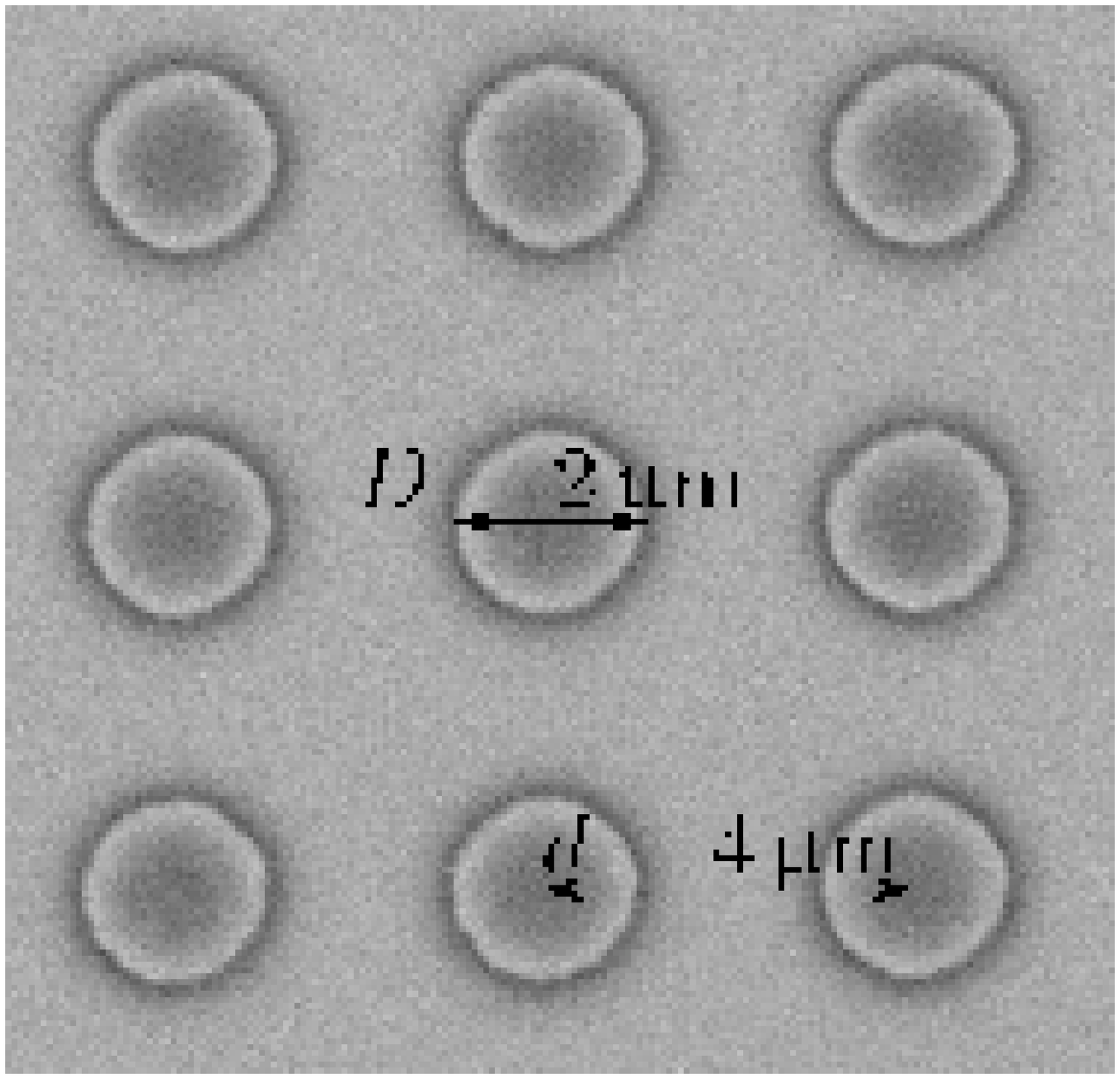}}
 \end{tabular}
\end{center}
\caption{Geometry of the studied structures. (a): schematic representation and notations. 
(b): SEM image (top view) of a fabricated grating.
\label{schema}}
\end{figure}
%
\subsection{Diffraction problem}\label{DP}
The geometry of the structures studied in this paper is represented in Fig.~\ref{schema1} and consist of three layers.
The top layer is made of a square array with period $d$ along both $Ox$ and $Oy$ 
of cylindrical chromium nanorods with diameter $D$ and thickness $h_{\rm r}$. 
The bottom layer is a continuous chromium film of thickness $h_{\rm m}$. These two metallic layers are separated by 
an amorphous silicon film of thickness denoted $h_{\rm i}$. The incident medium (superstrate) is air with permittivity $\varepsilon^+=1$ and 
the structure are deposited on a silicon substrate with permittivity $\varepsilon^-$. 
The permittivity of chromium is described by a Drude-Lorentz model \cite{Lovrin2009} and the refractive index of bulk and amorphous silicon 
are taken from tabulated data \cite{palik}. All materials are assumed to be non magnetic ($\mu_r=1$).\\
We consider here the time-harmonic regime with $\e^{-\ic\omega t}$ dependance. 
The structure is illuminated by a plane wave $\B E^{\mathrm{inc}} = \B{A}^{\mathrm{0}}\;\textrm{exp}(i \,\B{k}^+\cdot\B{r})$
 with 
 \begin{equation*}
    \B{k}^+ = \left| \begin{array}{l} \alpha\\\beta\\\gamma
    \end{array} \right. = k^+ \left| \begin{array}{l} -\sin \theta_0 \,\cos \varphi_0 \\-\sin \theta_0 \,\sin \varphi_0 \\-\cos \theta_0
    \end{array} \right.
\end{equation*}
and
\begin{equation*}
    \B{A}_0^e = \left| \begin{array}{l} E^0_x\\E^0_y\\E^0_z
    \end{array} \right. =A^e  \left| \begin{array}{l} \cos \psi_0 \, \cos \theta_0  \,\cos \varphi_0  - \sin \psi_0 \, \sin \varphi_0 \\\cos \psi_0 \, \cos \theta_0 \, \sin \varphi_0  + \sin \psi_0 \, \cos \varphi_0 \\-\cos \psi_0 \, \sin \theta_0
    \end{array} \right.
\end{equation*}
where $\varphi_0\in[0,2\pi]$, $\theta_0\in[0,\frac{\pi}{2}]$, $\psi_0\in[0,\pi]$, $k_0=\omega/c$ and $k^+=k_0\sqrt{\varepsilon^+}$.\\
The problem we are dealing with is to find non trivial solutions 
of Maxwell's equation, \textit{i.~e.} to find the unique electromagnetic field $(\B E,\B H)$ such that 
\begin{equation}
    \Lop_{\tens{\varepsilon},\tens{\mu}}(\B E):=-\rot\left(\tens{\mu}^{-1}\,\rot\B E\right) 
    + k_0^2\,\tens{\varepsilon}\,\B E = {\B 0} ,
    \label{helm3D}
\end{equation}
where the diffracted field $\B E^d=\B E-\B E^{\rm inc}$ satisfies an outgoing wave condition (OWC) and 
where $\B E$ is quasiperiodic along $x$ and $y$ 
\begin{equation*}
 \B E(x+d_x,y+d_y,z)=\B E(x,y,z)\e^{\ic(\alpha d_x+\beta d_y)}.
\end{equation*}
Under this form, the problem is not adapted to a resolution by a numerical method 
because of infinite issues: the sources of the plane wave are infinitely far above the structure, 
the geometric domain is unbounded and 
the scattering structure is itself infinitely periodic. 
To circumvent these issues, we compute only the 
diffracted field solution of an equivalent radiation problem with sources inside the scatterers, 
we use PMLs to truncate the unbounded domain at
a finite distance, and we use quasiperiodicity conditions to model a single period of the grating.\\
Denoting $\tens{\varepsilon_1}$ and $\tens{\mu_1}$ the tensor fields describing the
multilayer problem, the function $\B E_1$ is defined as the unique solution of
 $\Lop_{\tens{\varepsilon_1},\tens{\mu_1}}(\B E_1)=0$, such that $\B E_1^d:=\B E_1-\B E_0$ 
 satisfies an OWC. The expression of this function can be 
calculated with a matrix transfer formalism extensively used 
in thin film optics (See for example Ref. \cite{mcleod}). The unknown function $\B E_2^d$ is thus given by
 $\B E_2^d=\B E-\B E_1=\B E^d-\B E_1^d$.
 The scattering problem (\ref{helm3D}) can be rewritten as:
\begin{equation}
\Lop_{\tens{\varepsilon},\tens{\mu}}(\B E_2^d)=-\Lop_{\tens{\varepsilon_1},\tens{\mu_1}}(\B E_1):=\source_1.
\label{equ:u2d}
\end{equation}
The term on the right hand side can be seen as a source term $\source_1$ with support in 
the diffractive objects $\Omega_{g'}$ and is known in closed form \cite{Demesy2009}.\\
The radiation problem defined by
Eq.~(\ref{equ:u2d}) is then solved by the FEM \cite{Demesy2007,Demesy2009,Demesy2009a}, using PMLs to truncate the infinite regions and by
setting convenient boundary conditions on the outermost limits of the domain. 
We apply Bloch quasiperiodicity conditions with 
coefficient $\alpha$ (resp. $\beta$) on the two
parallel boundaries orthogonal to $x$ (resp. $y$), and homogeneous 
Dirichlet boundary conditions on the outward boundary of the PMLs. 
The computational cell is meshed using $2\textsuperscript{nd}$ order edge elements. 
The final algebraic system is solved using a direct solver (PARDISO \cite{Schenk2004475}).

\subsection{Spectral problem}\label{part:eigenpb}
The diffractive properties of open waveguides such as those studied here are governed by their eigenmodes and eigenfrequencies. 
The eigenproblem we are dealing with consists in finding the solutions of source free Maxwell's equations, \textit{i.e.}
 finding eigenvalues ${\Lambda_n}=({\omega}_n/c)^2$ and non zero eigenvectors $\B V_n$ such that:
\begin{equation}
\Mop_{\tens{\mu}}(\B V_n):=\rot\left(\tens{\mu}^{-1}\,\rot\B V_n\right) =\Lambda_n \,\tens{\varepsilon} \, \B V_n.
\label{eq:eigenpb}
\end{equation}
Note that we search for Bloch-Floquet eigenmodes so Maxwell's operator 
$\Mop_{\tens{\mu}}$ is parametrized by the real quasiperiodicity coefficients $\alpha$ and $\beta$. 
Because we are dealing with an open structure, the eigenvalues $\Lambda_n$ are complex even for Hermitian materials.  
The spectrum of the associated Maxwell's operator is constituted of a continuous part corresponding to radiation modes and a 
discrete set of complex eigenvalues associated with the so-called quasimodes (also known as leaky modes or resonant states). 
PMLs have proven to be a very convenient tool to 
compute leaky modes in various configurations \cite{pmlCompel,Popovic2003,HEIN2004,Eliseev2005} because they mimic
efficiently the infinite space provided a suitable choice of their parameters. Indeed, if we choose a constant 
stretching parameter $\zeta$ for the PMLs, it is sufficient to take $\re(\zeta)>0$ and $\im(\zeta)>0$ to rotate the continuous spectrum 
in the lower half complex plane $\re(\omega)<0$, which reveals \emph{outgoing} quasimodes (satisfying outgoing wave conditions) \cite{VialPRA2014}. It is well known that the
associated eigenvalues are \emph{poles} of the scattering matrix. In addition, the \emph{zeros} $\Lambda^z_n$ of the scattering matrix are associated with 
\emph{incoming} quasimodes (satisfying incoming wave conditions), that we can compute by setting $\re(\zeta)>0$ and $\im(\zeta)<0$ 
leading to a displacement of the continuous spectrum 
in the upper half complex plane $\re(\omega)>0$. A real zero $\Lambda^z$ indicates total absorption of incident light.\\
Note that the incident angles $\theta_0$ et $\varphi_0$ appear in a subtle way through the quasiperiodicity coefficients $\alpha$ et $\beta$, 
but the polarization angle $\psi_0$ \emph{does not come into play in the spectral problem}. It is thus necessary to thoroughly study eigenmodes 
in order to find the polarization state that can excite the modes at stake.\\

The eigenvalue problem defined by Eq.~(\ref{eq:eigenpb}) is solved with the FEM as described in section \ref{DP}.
We have supposed here that the material are non dispersive, which makes the problem in Eq.~(\ref{eq:eigenpb}) linear. 
To take into account dispersion, the eigenvalue problem is solved iteratively with updated values of permittivity. 
This procedure converges rapidly due to the slow variations of the permittivity of the considered materials in the 
far infrared range.

\subsection{Quasimodal expansion method}

We first define the classical inner product of two functions $\B F$ and $\B G$ of $L^2(\Omega)$, $\Omega\subset\mathbb{R}^3$:
\begin{equation}
  \bra  \B F\mb \B G\ket :=\int_{\Omega} \B F(\B r)\cdotp \conj{\B G(\B r)} \; \ddroit\B r.
\end{equation} 
Unlike self-adjoint problems, $\bra \tens{\varepsilon} \B V_n \mb \B V_m\ket \neq\delta_{nm}$, in other 
words the eigenmodes $\B V_n$ are not orthogonal with respect to this standard definition. This is the reason why we 
consider an adjoint spectral problem with eigenvalues $\conj{\Lambda_n}=(\conj{\omega_n}/c)^2$
and eigenvectors $\B W_n$. The adjoint operator $\Mop^\dagger_{\tens{\mu}}$ is defined by
\begin{equation}
\bra \vphantom{ \Mop^\dagger_{\tens{\mu}}(\B W)}\Mop_{\tens{\mu}}(\B V) \mb \B W \ket 
=\bra  \B V \mb \Mop^\dagger_{\tens{\mu}}(\B W)\ket 
\end{equation}
with \emph{complex conjugate coefficients} for the boundary conditions in comparison with 
the direct spectral problem \footnote{actually, the boundary conditions employed here are identical for both spectral problems 
since we use only real valued coefficents (homogeneous Neumann boundary condition and real quasiperiodicity constants $\alpha$ and $\beta$).}, 
and is such that $\Mop^\dagger_{\tens{\mu}}=\Mop_{\tens{\mu}^\star}$, 
where $A^\star={\conj{A}}^\mathrm{T}$ is the conjugate transpose of  matrix $A$.
The associated adjoint problem that we shall solve is:
\begin{equation}
\Mop^\dagger_{\tens{\mu}}(\B W_n)=\rot\left({\tens{\mu}^\star}^{-1}\,\rot\B W_n\right) = 
\conj{\Lambda_n}\,{\tens{\varepsilon^\star}}\, \B W_n.
\label{eq:eigenpb_adjoint}
\end{equation}
We know from spectral theory that the eigenvectors $\B V_n$ are bi-orthogonal to their adjoint 
counterparts $\B W_n$ \cite{hanson2002operator}:
\begin{equation}
 \bra  \tens{\varepsilon} \B V_n \mb \B W_m\ket 
 :=\int_{\Omega}\tens{\varepsilon}(\B r)\, \B V_n(\B r)\cdotp \conj{\B W_m(\B r)} \; \ddroit\B r
 =K_n\delta_{nm},
\label{eq:biortho}
\end{equation} 
where the normalization coefficient $K_n= \bra\tens{\varepsilon}  \B V_n \mb \B W_n \ket$. 
Relation (\ref{eq:biortho}) provides a complete bi-orthogonal set to expand every field solution
of Eq.~(\ref{equ:u2d}) propagating in the open waveguide as:
\begin{eqnarray}
\displaystyle
  \B E_2^d(\B r,\omega,\psi)&=&\sum_{n=1}^{+\infty}P_n(\omega,\psi)\, \B V_n(\B r) \nonumber\\
  &&+ \int_{\Gamma_c} P_\nu(\omega,\psi)\, \B V_\nu(\B r) \; \ddroit\nu,
\label{decompmod_integrale}
\end{eqnarray}
where $\Gamma_c$ is the continuous spectrum (a curve, with possibly a denombrable set of branches in the complex plane). 
The coefficients $P_k(\omega,\psi)$, $k=\{n,\nu\}$, are given by:
\begin{equation}
 P_k(\omega,\psi)=\frac{1}{K_k}\bra  \tens{\varepsilon} \B E_2^d  \mb  \B W_k\ket =\frac{J_k(\omega,\psi)}{{\omega}^2-\omega_k^2},
 \label{equPn}
\end{equation}
with
\begin{eqnarray}
J_k(\omega,\psi)&=&\frac{c^2}{K_k}\bra \source_1  \mb   \B W_k \ket \nonumber\\
&=&\frac{c^2}{K_k}\int_{\Omega_{g'}}\source_1(\B r,\omega,\psi)\, \conj{\B W_k(\B r)} \;\ddroit\B r,
\label{equJn}
\end{eqnarray}
where the integration is \emph{only performed on the inhomogeneities} $\Omega_{g'}$ since the 
source term $\source_1$ is zero elsewhere. Note that the last integral has to be taken in the distributional meaning 
which leads to a surface term on $\partial\Omega_{g'}$ because of the spatial derivatives in $\source_1$.\\
We are thus able to know how a given mode is excited when changing the incident field.
This modal expansion can be approximated by a discrete sum since the spectrum of the final
operator we solve for
involves only discrete eigenfrequencies, and in practice only a finite number $M$ of modes is 
retained in the expansion, so that we can write:
\begin{equation}
 \B E_2^d(\B r,\omega,\psi)\simeq\sum_{m=1}^{M}P_m(\omega,\psi)\, \B V_m(\B r).
 \label{decompmod_rom}
\end{equation}
This leads to a reduced modal representation of the field which is well adapted when studying
the resonant properties of the open structure, as illustrated in the sequel.\\

\section{Modal analysis of MIM arrays}

The parameters employed are $h_{\rm r}=\SI{100}{nm}$, $h_{\rm i}=\SI{530}{nm}$, $h_{\rm m}=\SI{200}{nm}$, and 
we fix the ratio between the rod diameter and the period $f=D/d=0.5$. 
We study the influence of the period $d$ on the reflection spectrum of the metamaterial.\\

\subsection{Fabrication and characterization of the samples}
Samples with parameters described above and varying period of $4.0$, $4.4$, $4.8$, $5.2$ and $\SI{5.6}{\micro\meter}$ 
have been fabricated (a SEM image showing a top view of the filter with $d=\SI{4}{\micro\meter}$ is given in Fig~\ref{fab}).
 The different layers have been deposited by magnetron sputtering on a standard silicon 
wafer of diameter $\SI{100}{\milli\meter}$ and thickness $\SI{525}{\micro\meter}$. Large area samples
($\SI{1}{\centi\meter}\times\SI{1}{\centi\meter}$) were patterned with a standard photolithography process 
with a positive resist deposition followed by a chemical etching of the top chromium layer.\\
Reflection spectra have been recorded with a Thermo Fisher-Nicolet 6700 Fourier Transform InfraRed (FTIR) spectrophotometer. 
The measurements were performed with a focused unpolarized light beam with $\pm \SI{16}{\degree}$ divergence and a 
spot diameter of $\SI{4}{\milli\meter}$. An accessory composed of a set of mirrors allows us to record reflection spectrum 
for incident angles between $0$ and $\SI{90}{\degree}$. All the spectra are normalized with a background recorded 
from a reference gold mirror.

\subsection{Reflection spectra}

\begin{figure}[htbp!]
\begin{center}
\begin{tabular}{c}
\subfigure[~Simulations.\label{spectresRsim}]{\includegraphics[width=0.95\columnwidth]{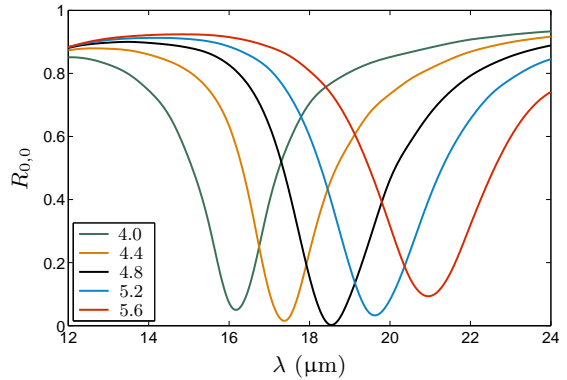}}\\
\subfigure[~Experiments.\label{spectresRexp}]{\includegraphics[width=0.95\columnwidth]{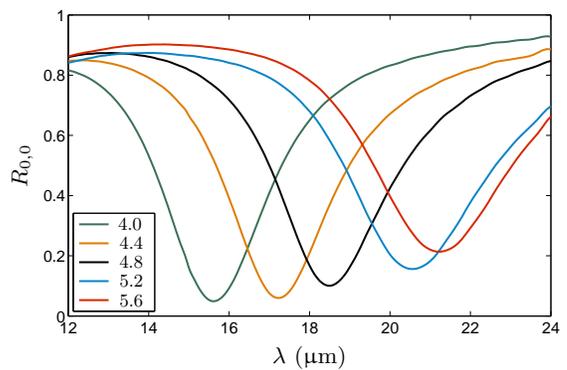}}
 \end{tabular}
\end{center}
\caption{Reflexion spectrum at normal incidence in the specular order $R_{0,0}$ as a function of incident wavelength $\lambda$ 
for different values of the period $d$ (in $\si{\micro\meter}$). 
(a): FEM simulations, (b): FTIR measurements.}
\label{spectresR}
\end{figure}
Figure~\ref{spectresRsim} shows the reflection spectra at normal incidence in the specular order for bi-gratings with different periods, calculated 
by the FEM formulation described in section~\ref{DP}. These spectra show
a clear resonant behavior in the region $12-\SI{24}{\micro\meter}$ with a large reflection dip. 
Increasing the period $d$ shifts this dip to larger wavelengths and broadens the resonance. 
It can also be noted that for $d=\SI{4.8}{\micro\meter}$, the reflection is almost zero at resonance. Since 
the transmission is negligible because the thickness of the bottom metal layer is nearly twice the skin depth of chromium in this 
spectral range, the incident power is nearly totally absorbed by the metamaterial at resonance and dissipated by Joule heating.\\
The measured reflection spectra of the fabricated samples are reported in Fig.~\ref{spectresRexp} and show very good agreement with 
numerical simulations. For example for $d=\SI{4.8}{\micro\meter}$, both experimental and simulated reflection dips 
are located at $\SI{18.5}{\micro\meter}$, although experimentally, the reflection minimum is 10\%, more than the 0.3\% simulated value. 
For all samples, the disagreements originates from spectral broadening of the measured reflection, 
wich is mainly due to size dispersion on the rod diameter over the fabricated samples.

\subsection{Influence of the periodicity: a pole-zero approach}

To highlight the resonant properties of the studied MIM arrays, we report here a modal analysis of such structures. 
We solved numerically the spectral problem (\ref{eq:eigenpb}) as described in section \ref{part:eigenpb}, with 
quasiperiodicity coefficients $\alpha=\beta=0$. Due to the symmetry of the problem in these conditions, we find two degenerate outgoing leaky modes 
(associated with poles of the complex reflection coefficient $r_{0,0}$) and two degenerate incoming leaky modes (associated with zeros of $r_{0,0}$). 
The degenerescence corresponds to eigenmodes with TE and TM polarization.\\

\begin{figure}[htbp!]
\includegraphics[width=0.95\columnwidth]{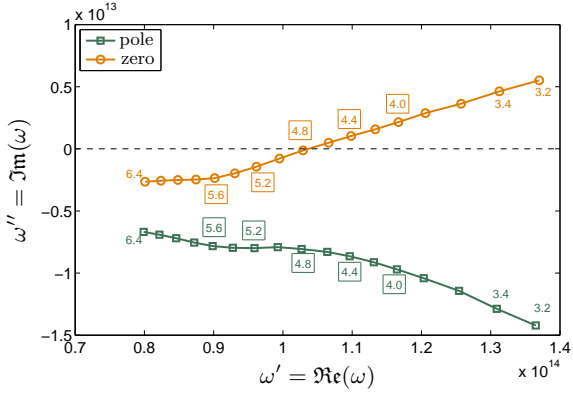}
\caption{Location of pole (green squares) and zero (orange circles) in the complex plane as a function of $d$ 
(we only represented the pole and zero of the TE mode because of degeneracy). 
The values of $d$ are indicated in $\si{\micro\meter}$, and the boxed values 
indicates the fabricated structures. The black dashed line represents the real axis.\label{planC}}
\end{figure}

Figure~\ref{planC} shows the evolution of the pole and its associated zero 
 in the complex $\omega$-plane as a function of $d$ (we only represented the pole and zero of the TE mode because of degeneracy). 
 The real parts of the pole and of the zero are almost equal and shift to smaller frequencies as the period increases. 
For $d=\SI{4.8}{\micro\meter}$, the zero crosses the real axis, which means that the reflection is suppressed 
for a real incident frequency close to this zero. This is consistent with the previous observations from reflection spectra.\\

\begin{figure}[htbp!]
\begin{center}
\begin{tabular}{c}
\subfigure[~Resonant wavelength.\label{params_spec_lamb}]{\includegraphics[width=0.95\columnwidth]{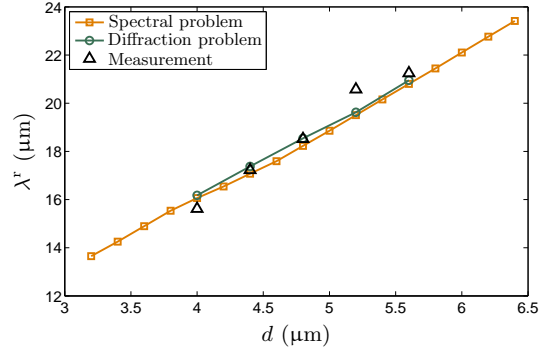}}\\
\subfigure[~Spectral width.\label{params_spec_larg}]{\includegraphics[width=0.95\columnwidth]{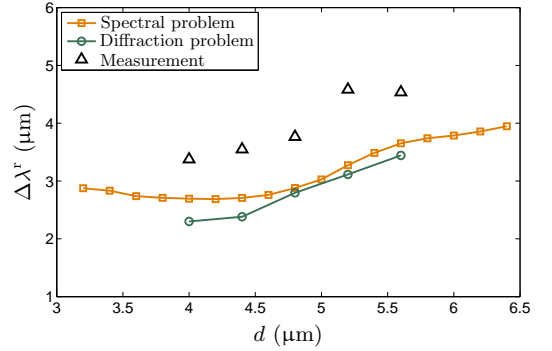}}
 \end{tabular}
\end{center}
\caption{Spectral parameters of the resonance as a function of the period $d$ obtained with different methods: 
extracted from calculated (green circles) and measured (black triangles) reflection spectra and extracted from the pole eigenfrequency (orange squares).
(a): resonant wavelength, (b): spectral width.}
\label{params_spec}
\end{figure}

We reported in Fig.~\ref{params_spec} the values of the resonant wavelength 
and the spectral width of the resonances extracted from the calculated (green circles) and from measured (black triangles) reflection spectra 
as well as those derived from the pole eigenfrequencies (orange squares). As it can be seen 
in Fig.~\ref{params_spec_lamb}, the position of the resonance increases linearly with $d$, with the
values calculated from simulated reflection spectra minima and from the spectral problem being in excellent agreement, which 
indicates that the resonant reflection dip stems from the excitation of the 
leaky mode associated with this eigenfrequency. In addition, experimental values well agree with 
the positions predicted by the two numerical approaches.
Moreover, the spectral width of the dip increases with $d$ as can bee seen in Fig.~\ref{params_spec_larg}. 
In that case the values obtained from the diffraction problem and from the spectral problem are in good agreement but slightly differs 
because the spectral width extracted from reflection spectra may be influenced by the presence of other modes whereas the 
linewidth associated with a leaky mode is valid for an isolated resonance. The experimental values are larger as said before but show a 
variation with $d$ similar to the calculated ones.\\

\begin{figure}[htbp!]
\includegraphics[width=0.95\columnwidth]{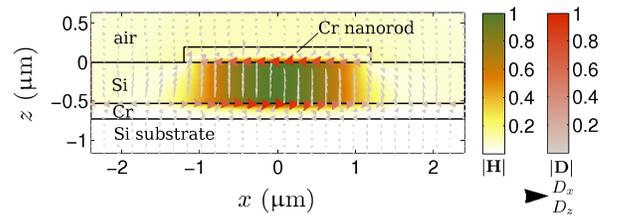}
\caption{Field map of the TE outgoing leaky mode in the $Oxz$ plane for $d=\SI{4.8}{\micro\meter}$. The left colormap 
represents the norm of the magnetic field $\B H$, the arrows represent the direction of the 
displacement current $\B D$ and their colors (right colormap) and size are proportional to its intensity.}
\label{mode}
\end{figure}

To highlight the physical mechanism responsible for these resonant total absorption (or equivalently suppressed reflection), 
we plotted in Fig.~\ref{mode} the magnetic field associated with the TE outgoing quasimode for $d=\SI{4.8}{\micro\meter}$. 
The electric displacement represented by arrows is very strong with opposite directions in the rod and the metal layer, 
which creates a strong magnetic response (see colormap) confined in the silicon layer below the nanorod. 
Note that the nature of the resonance is not related to Fabry-P\'erot type mechanism because the 
silicon layer is very thin ($<\lambda/30$), but rather to localized electric and magnetic dipoles \cite{Hao2010}. 

\subsection{Angular tolerance}

\begin{figure*}[htbp!]
\begin{center}
\begin{tabular}{cc}
\subfigure[~Simulations, TE.\label{spectresRfnbetate}]{\includegraphics[width=0.95\columnwidth]{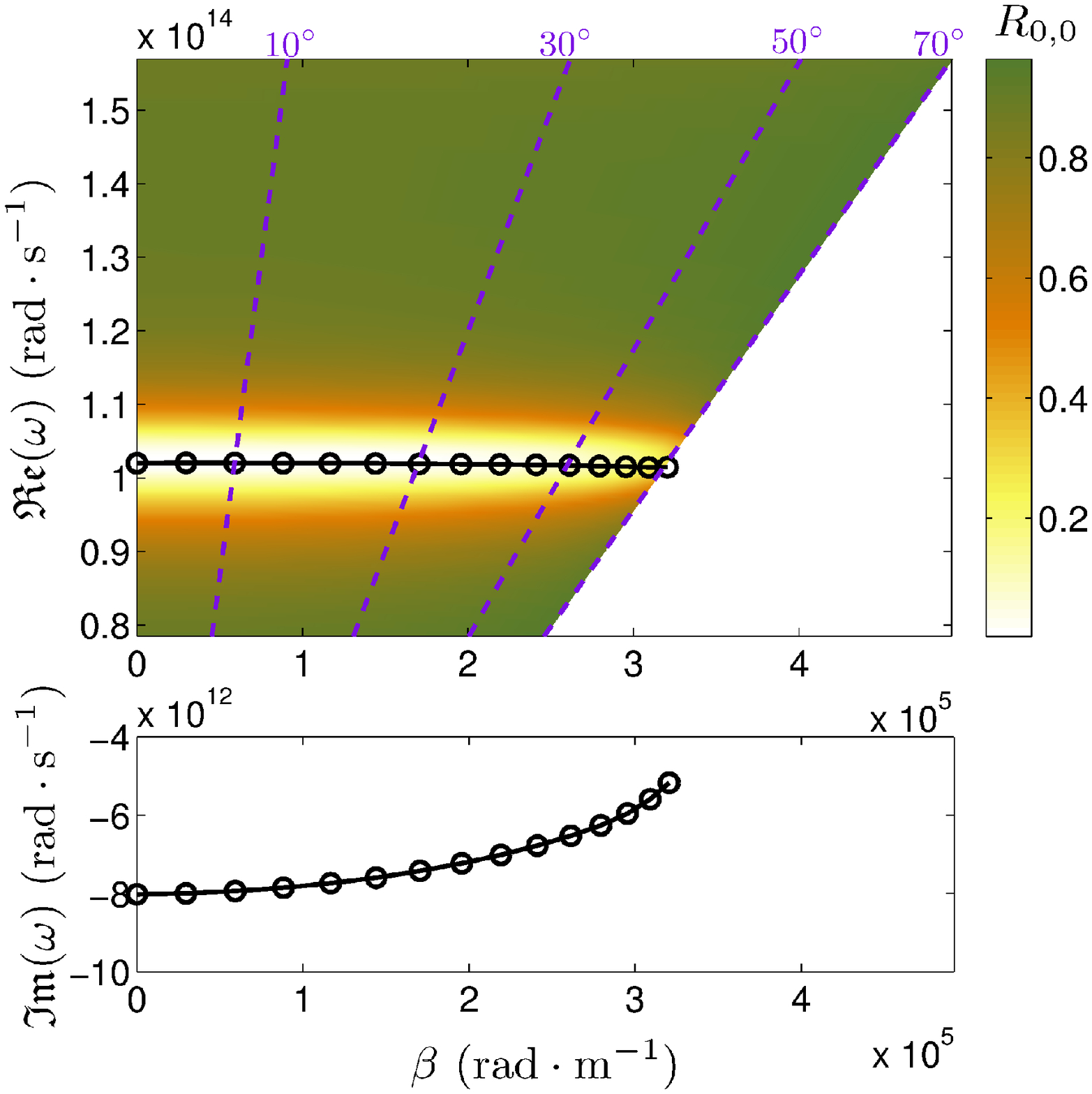}}
\subfigure[~Simulations, TM.\label{spectresRfnbetatm}]{\includegraphics[width=0.95\columnwidth]{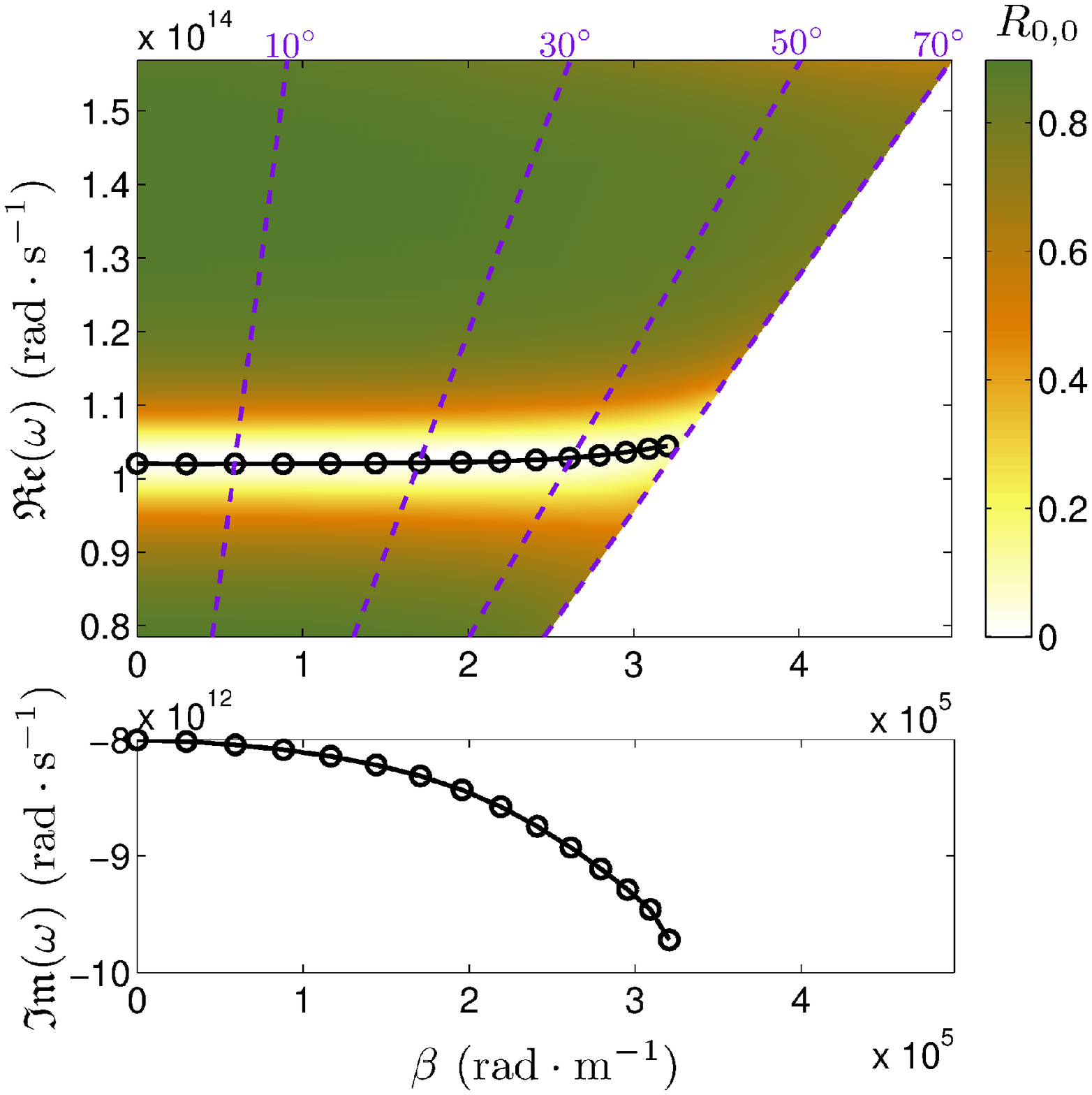}}\\
\subfigure[~Simulations, unpolarized.\label{spectresRfnbetaup}]{\includegraphics[width=0.95\columnwidth]{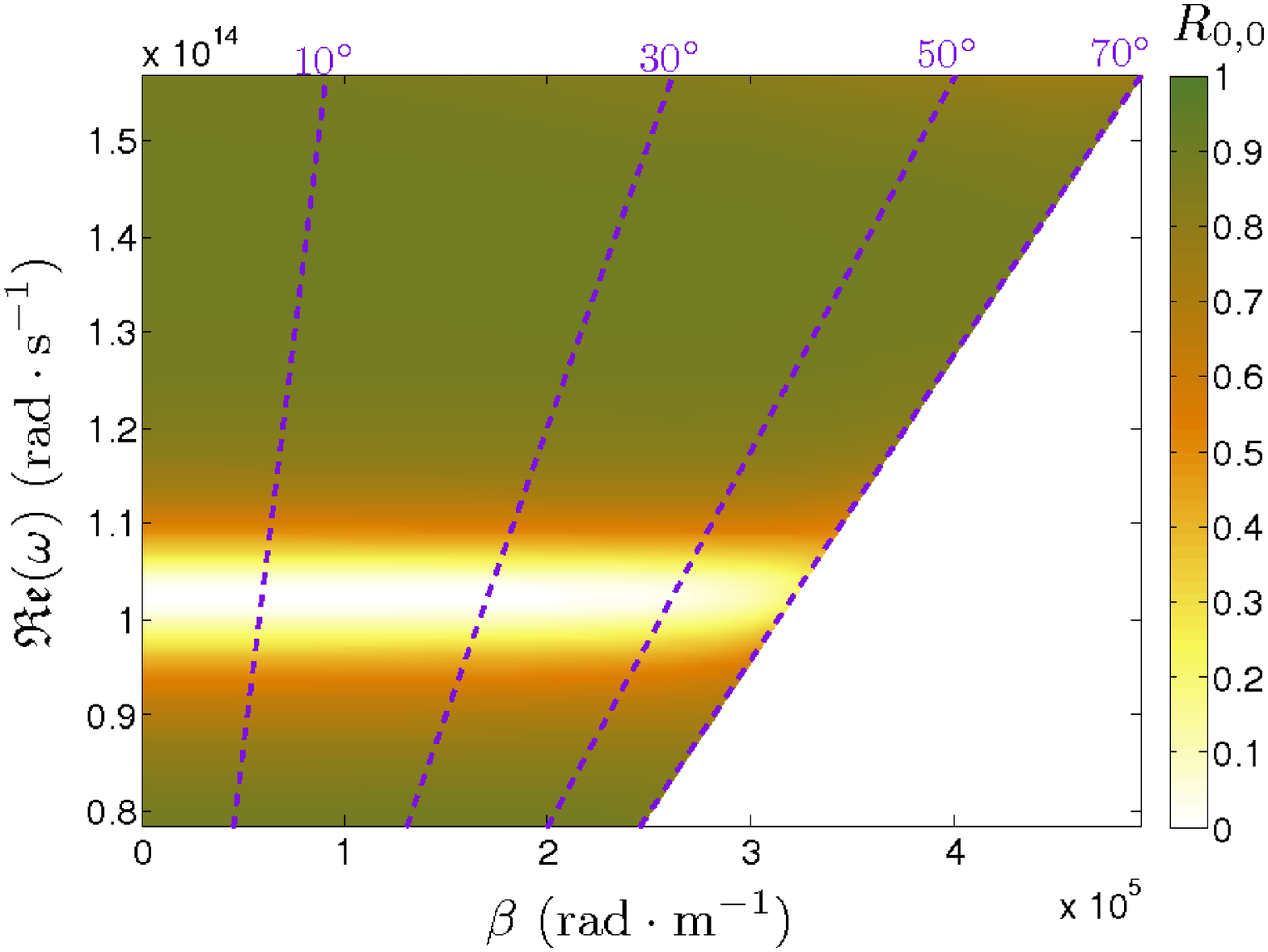}}
\subfigure[~Experiments.\label{spectresRfnbetames}]{\includegraphics[width=0.95\columnwidth]{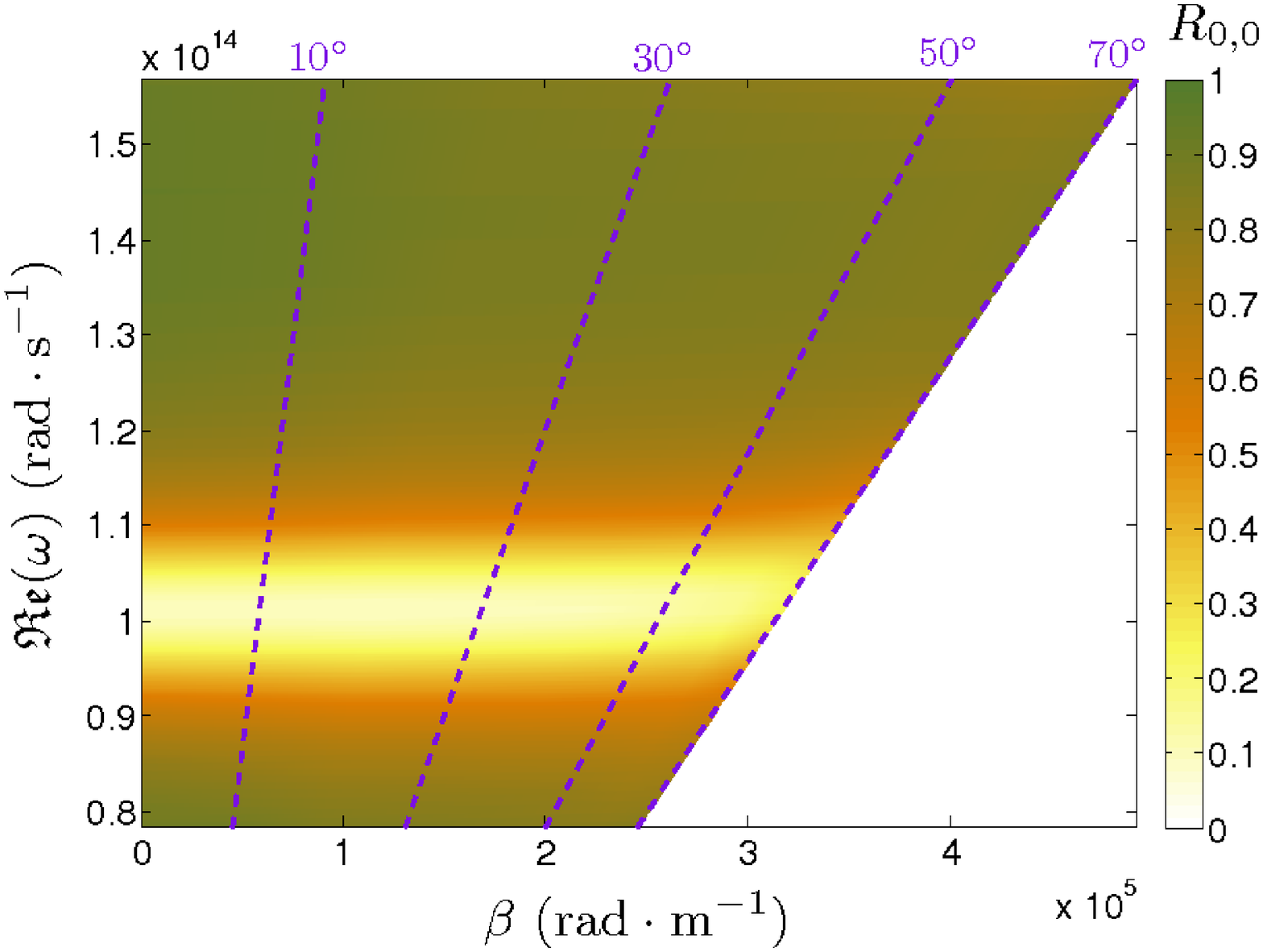}}
 \end{tabular}
\end{center}
\caption{Influence of the incidence. Colormap : reflection spectrum in the specular order $R_{0,0}$ as a function of 
frequency $\omega$ and quasiperiodicity coefficient $\beta$ for $d=\SI{4.8}{\micro\meter}$. (a): simulations TE polarization, (b): simulations TM polarization , 
(c): simulations unpolarized, (d): FTIR measurements. In Figs. (a) and (b), the black circles indicate 
the real (top) and  imaginary (bottom) parts of the eigenfrequency $\omega_1$ of the corresponding leaky mode as a function of $\beta$.}
\label{spectresRfnbeta}
\end{figure*}

One of the key features of MIM arrays is the angular tolerance of the first order 
resonance, which is crucial for filtering applications. The colormap on Fig~\ref{spectresRfnbeta} shows the 
reflection spectrum as function of frequency $\omega$ and transverse wavenumber $\beta$ for $d=\SI{4.8}{\micro\meter}$. 
Figs.~\ref{spectresRfnbetate} and~\ref{spectresRfnbetatm} are calculated values in TE and TM polarization respectively. 
We also plotted the evolution of the real part $\omega'_1$ of the eigenfrequency $\omega_1$ associated with either TE or TM mode, 
the so-called dispersion diagram. In both cases the real part of the eigenfrequency remains almost constant, with a slight redshift (resp. blueshift) 
for TM (resp. TE) polarization at large angles and matches very well the position of the resonant reflection dip. As $\beta$ 
increases, the resonance sharpens in the TE case and broadens in the TM case. These observations are confirmed 
by the evolution of imaginary part $\omega''_1$ of eigenfrequencies (See bottom plot in Figs.~\ref{spectresRfnbetate} and  \ref{spectresRfnbetatm}): 
because the real part $\omega'$ is almost constant the quality factor of the resonance $Q=\omega'/\Delta\omega=\omega'_1/2\omega''_1$ 
increases (resp. decreases) for TE (resp. TM) polarization. To compare with experimental results of Fig.~\ref{spectresRfnbetames}, we also 
plotted the calculated unpolarized case in Fig.~\ref{spectresRfnbetaup}. The agreement between 
simulations and measurements is excellent except a slight spectral broadening and higher minimum values for experimental results and 
demonstrates the angular tolerance up to $\SI{70}{\degree}$ of the fabricated filters.

\subsection{Leaky mode excitation and reduced order model}

\begin{figure}[t!]
\begin{center}
\begin{tabular}{cc}
\subfigure[~$\psi=0$ (TE)]{\includegraphics[width=0.9\linewidth]{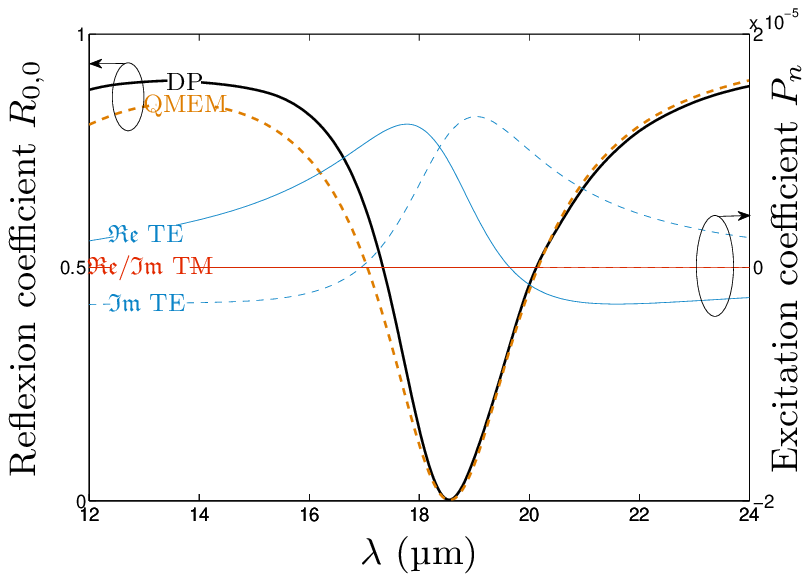}}\\
\subfigure[~$\psi=\pi/2$ (TM)]{\includegraphics[width=0.9\linewidth]{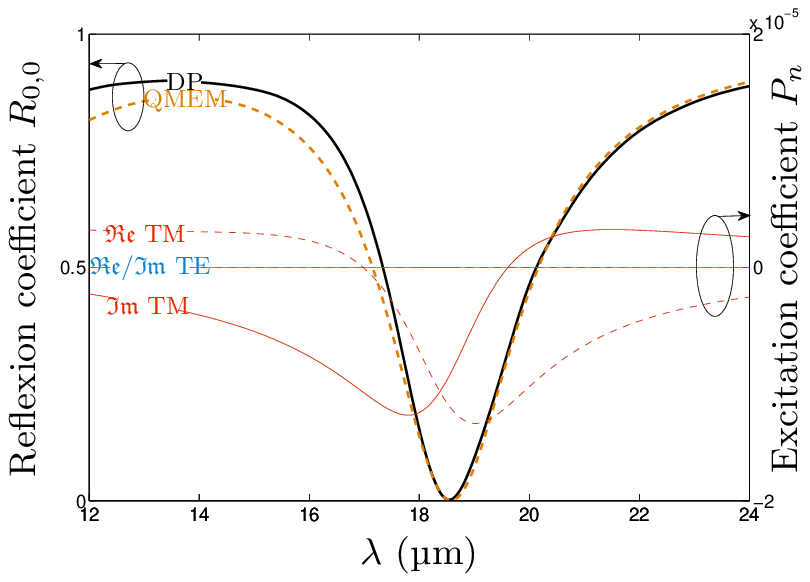}}\\
\subfigure[~$\psi=\pi/4$]{\includegraphics[width=0.9\linewidth]{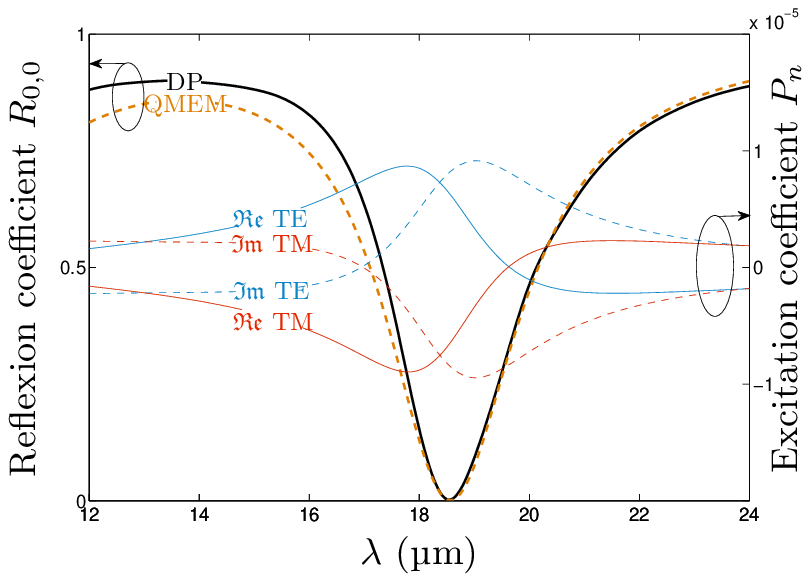}}
 \end{tabular}
\end{center}
\caption{Excitation coefficients $P_n$ for the two degenerate modes (right ordinate, TE blue, TM red, real parts: solid line, imaginary part: dashed line) and 
reflection coefficient $R_{0,0}$ computed with full wave FEM diffraction problem (DP, black solid line) and 
with the QMEM with these two modes (orange dashed line).
}
\label{Rreconstr_Pn}
\end{figure}
Finally, we computed the diffracted field using  Eq.~(\ref{decompmod_rom}) with the two leaky modes TE and TM. 
Because of the mode degeneracy, every linear combination of the two eigenmodes is also solution of Eq.~(\ref{eq:eigenpb}) 
for the eigenvalue denoted $\omega_1=\omega'_1+\ic \omega''_1$. We define the TE mode such that 
$J_{\rm{TE}}(\omega_1',\psi_{\rm{TE}})=1$
 and $J_{\rm{TE}}(\omega_1',\psi_{\rm{TM}})=0$, where 
 $\psi_{\rm{TE}}=0$ and $\psi_{\rm{TM}}=\pi/2$. The TM mode is then obtained by standard Gram-Schmidt 
 orthogonalization procedure, and the two modes are finally normalized such that $K_{\rm{TE}}=K_{\rm{TM}}=1$. \\

The study of the coupling coefficients $P_n$ reveals the resonant nature of the interaction of a plane wave 
with the modes. On Fig.~\ref{Rreconstr_Pn}, we plot these coefficients as a function of 
wavelength for different polarization cases: (a) $\psi=0$ (TE), (b) $\psi=\pi/2$ (TM) and 
(c) $\psi=\pi/4$. The real (solid line) and imaginary (dashed line) parts of the excitation coefficients show strong 
variations around the resonant frequency in all cases. For $\psi=0$ (resp. $\psi=\pi/2$), only the 
TE (resp. TM) mode is excited while the value of $P_n$ for the TM (resp. TE) mode is negligible. 
For $\psi=\pi/4$ both modes participate equally to the resonant diffraction process as 
their coupling coefficients are equal in absolute value (opposite sign is arbitrarily set for display purpose). 
These observations illustrate the independence of the reflection dip with regards to polarization. We have also computed
 $R_{0,0}$ with the field reconstructed by the QMEM 
 with only two leaky modes. The results (orange dashed line on Fig.~\ref{Rreconstr_Pn}) 
 are in all cases in excellent agreement with full wave FEM simulations of the diffraction problem
 (DP, black solid line). This means that the diffractive properties of the structure are dominated by these two modes 
 in the considered wavelength range. The small discrepancies at small wavelengths are attributed to other modes 
 with higher resonant frequencies not taken into account in the reduced order model.
 
\section{Conclusion}
We have studied metamaterial based on MIM designed to serve as reflection bandcut filters 
in the thermal infrared spectral range. These structures shows quasi total absorption of light at the resonant wavelength 
that can be tuned by varying the lateral dimensions of the metallic nanorods grating. The reflection dip spectral position 
is also independent of incident angle up to $\SI{70}{\degree}$ and is not affected by the polarization state of the incident light. 
Our study provides an in depth modal analysis revealing the 
resonant nature of the interaction of light with leaky modes of the structure. We developed a 
quasimodal expansion method (QMEM) that allows us to compute coupling coefficients between 
a plane wave and the modes. This method leads to a reduced order model with two modes that fits very well 
full wave FEM diffraction problem simulations. Large area samples have been fabricated 
and FTIR measured reflection spectra are in good agreement 
with the different numerical approaches, demonstrating the potential practical application 
of those polarization independent and angular tolerant resonant filters. Although the filters 
studied here have been designed to work between $15$ and $\SI{22}{\micro\meter}$, 
the concepts studied here can be applied to higher frequency ranges (\textit{e.g.} band III of the infrared between $7$ and $\SI{13}{\micro\meter}$) 
by scaling down the dimensions of the structures.

\begin{acknowledgments}
This research was financially supported by the Fonds Unique Interministériel (FUI) and by a CIFRE fellowship from the 
french Agence Nationale de la Recherche et de la Technologie (ANRT).\\
Part of the components were realized within the framework of the Espace Photonique facility with the financial support of the French 
Department of Industry, the local administration (Provence-Alpes Côte d'Azur Regional Council), 
CNRS and the European Community.
\end{acknowledgments}

\bibliography{biblio_these.bib}

\end{document}